\documentclass[final,3p,times,twocolumn,sort&compress,10pt]{elsarticle}
\usepackage{lineno,hyperref,amssymb,amsmath}
\usepackage{graphicx,multicol}
\usepackage{float}
\usepackage{array}
\usepackage{caption}
\usepackage{subcaption}
\newcolumntype{C}[1]{>{\centering\arraybackslash}p{#1}}

\journal{Nuclear Instruments and Methods in Physics Research Section A: Accelerators, Spectrometers, Detectors and Associated Equipment}
\begin{document}
\makeatletter
\def\ps@pprintTitle{%
  \let\@oddhead\@empty
  \let\@evenhead\@empty
  \def\@oddfoot{\reset@font\hfil\thepage\hfil}
  \let\@evenfoot\@oddfoot
}
\makeatother

\begin{frontmatter}

\title{On the possibility of creating a UCN source on a periodic pulsed reactor}

\author[Institute1]{A.I. Frank\corref{mycorrespondingauthor}}
\ead{frank@nf.jinr.ru}

\author[Institute1]{G.V. Kulin\corref{mycorrespondingauthor}}
\cortext[mycorrespondingauthor]{Corresponding authors}
\ead{kulin@nf.jinr.ru}

\author[Institute1]{N.V. Rebrova}
\author[Institute1]{M.A. Zakharov}

\address[Institute1]{Joint Institute for Nuclear Research, Dubna, Russia}

\begin{abstract}

The possibility of creating a UCN source on a periodic pulsed reactor is considered. It is shown that the implementation of the time focusing principle based on non-stationary neutron diffraction and the idea of the UCN trap pulse filling allow creating a sufficiently intense UCN source in a moderate-power pulsed reactor.
\end{abstract}


\end{frontmatter}

\section{Introduction}

As it is known, the credit for the discovery of ultracold neutrons (UCNs) belongs to F. L. Shapiro and his collaborators~\cite{Luschikov1969}. In an experiment performed at a reactor with an average power of 6 kW, they observed neutrons capable of accumulating in a closed volume – a curved tube –for several seconds. For the history of this discovery, see~\cite{Strelkov2015}. In the world literature, Dubna is rightfully considered the birthplace of UCNs.

In recent years, several UCN sources have appeared in the world~\cite{Bison2017, Steyerl2020}, and several more are now under construction. Unfortunately, there is no UCN source in Dubna, although attempts have been made to create one~\cite{Ananiev1989}. The reasons for this are largely due to the features of the pulsed reactor at JINR~\cite{Ananiev1977, IMFrank2018, Aksenov2009}. Its average power of 2 MW is relatively low to create a continuous UCN source, and the 5-Hz repetition rate is too high to accumulate neutrons generated in a single pulse. At the same time, the pulse flux of thermal neutrons is exceptionally high, since the interval between the pulses significantly exceeds their duration, which is about 350 microseconds. Obviously, under certain conditions, the pulsed flux of UCNs from a thin moderator can be rather significant. Thus, the question is how to take advantage of this circumstance.
A possible solution to this problem was proposed by F. L. Shapiro~\cite{Shapiro1971, Shapiro1976}. It consists in filling the UCN trap only during the pulse and effectively isolating it for the rest of the time. In the ideal case of no loss, the UCN density in the trap will correspond to the peak neutron density, which may be several orders of magnitude higher than the time average.

The implementation of this idea is complicated by the fact that in practice the trap is at a certain distance from the moderator due to the presence of biological shielding. Therefore, it becomes necessary to create a several meters long transport neutron guide feeding the trap. Placing the isolation valve next to the moderator – the UCN source – causes the neutron guide to become part of the trap. Due to the small transverse size of the neutron guide, the frequency of neutron collisions with its walls is quite high. This significantly reduces the storage time of UCNs in the trap - neutron guide system and, accordingly, considerably reduces the gain. Placing the valve at the entrance to the trap, several meters away from the source, is useful only in the case of sources with a low repetition rate ~\cite{Anghel2009, Saunders2013, Lauss2014}. For sources with a repetition rate of several hertz, the spread of the UCN transit times will exceed the intervals between pulses, and the presence of a valve at the entrance to the trap will not make sense. This difficulty can be overcome by using a device that acts as a time lens and forms a time image of the source in the immediate vicinity of the trap. In this case, the pulse duration of the time image can be of the same order of magnitude as the duration of the true pulse flow of the UCNs in the source. A time lens can play almost the same role as an optical lighter in a conventional microscope. It permits matching the object under study with the image of the light source, while the source itself (lamp) cannot be matched with the object. 

This proposal was formulated in~\cite{AIFrank1996, AIFrank2000}, where the principles of forming a time image of a point source were discussed. Note that in the literature, the term "time lens" is used in relation to devices with significantly different functions. While the works of Frank and Gähler~\cite{AIFrank1996, AIFrank2000} dealt with a lens forming a given time distribution of the neutron flux at a known point in space, the works of Rauch and co-authors~\cite{Rauch1985, Summhammer1986} and the following work~\cite{Baumann2005} discussed the possibility of compressing the velocity interval at the observation point, that is, additional monochromatization of the neutron beam.
In this paper, we intend to show that the time-focusing principle can serve as a methodological basis for creation of intensive UCN sources in periodic acting pulsed reactors.

\section{Operation principle of a source with time focusing}
\subsection{Time lens and methods for changing the neutron energy}
The main element of the UCN source discussed here is a time lens, the operation of which is based on controlling the velocity and, accordingly, the energy of the neutron. Apparently, there is a fairly wide range of possibilities to do this. In works [15-17], the effect on the neutron movement was proposed to be provided due to the appropriately formed configuration of a stationary or alternating magnetic field. The change in the velocity of neutrons is achieved due to a force affecting them, $\mathbf{F}=\nabla(\mu \mathbf{B})$, where $\mu$ is a neutron magnetic moment, and B is magnetic induction. In this case, a magnetic time lens that forms a magnetic field is a device along the direction of the neutron beam.

\begin{figure*}[h]
\centering
\captionsetup[subfigure]{labelformat=empty}
\begin{subfigure}[b]{\columnwidth}
\caption{}
        \includegraphics[width=0.8\textwidth]{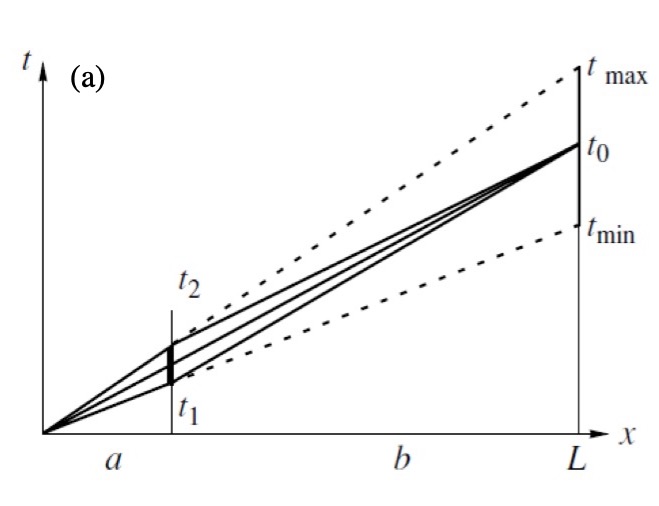}
        \label{subfig:Fig1a}
    \end{subfigure}
    \begin{subfigure}[b]{\columnwidth}
     \caption{} 
        \includegraphics[width=\textwidth]{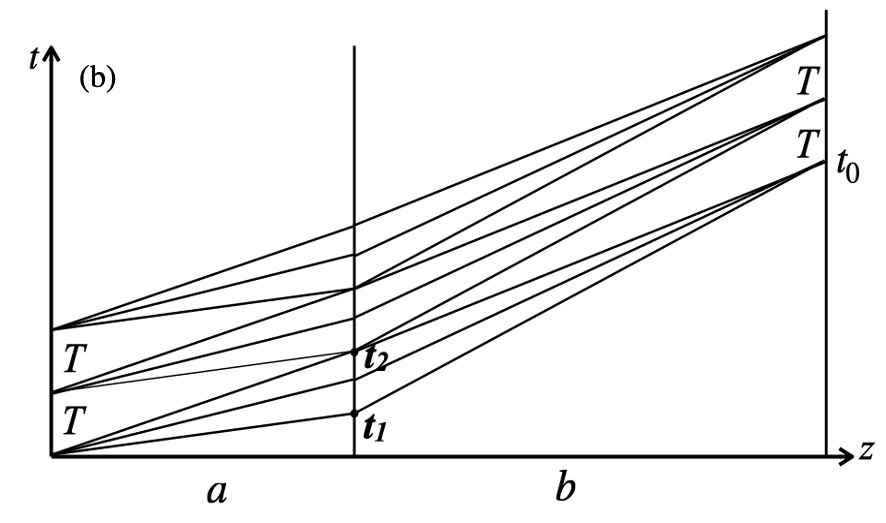}
        \label{subfig:Fig1b}
         \vspace{-0.7\baselineskip}
    \end{subfigure}
\caption
{
Scheme of the time lens action operation. (a) Lens that does not change the average neutron energy, (b) Lens that slows down the neutrons.
}
\label{fig:Fig1}
\end{figure*}

In contrast to the above-stated proposals, in~\cite{AIFrank1996, AIFrank2000} it was assumed that the lens is a local device. Let us explain the principle of its operation (see Fig.~\ref{subfig:Fig1a}).

Let us assume that at a time moment $t=0$ from a point $x=0$ neutrons are emitted in the positive direction of the $X$-axis. Their velocities $v$ are distributed in a certain range of values. The time of their arrival at the observation point $t_L$ will be distributed in the interval $t_1<t_L<t_2$. Supposedly, at a point $x=a$ a time lens is located, and it can change the neutron energy by the value $\Delta E(t)$, according to a given time law in the time interval $t_1<t_L<t_2$. The dependence $\Delta E(t)$ can be chosen in such a way that the velocities of neutrons that have passed through the lens fulfill the condition of simultaneous arrival at the observation point at the time $t_L=t_0$.

\begin{equation}\label{eq1}
\frac{a}{V_a} +\frac{b}{V_b}=t_0,\;\;\; a+b=L,
\end{equation}
where $V_a$ and $V_b$ are neutron velocities before and after passing the lens, respectively.

\begin{equation}\label{eq2}
\Delta E=\frac{m}{2} \left | \left(\frac{b}{t_0-t}\right)^2-\left(\frac{a}{t}\right)^2 \right |,\;\;\; t=\frac{a}{V_a},\;\;\; t_1<t<t_2,
\end{equation}
where $m$ is the neutron mass.

In the same way that image formation in optics is associated with transformation of the angular distribution of the beam, time focusing involves a change in the distribution of neutron velocities. At the focusing in time the duration of the time pulse is transformed, which allows us to introduce the concept of time magnification $M$.

Let us make the period of the lens action $T=t_2-t_1$ coincide with the period of the source pulses repetition. Thus, the repetition frequency $F_s=T^{-1}$, the distance from the source to the lens $a$, and the range of received neutron velocities $V_{a\;min}\leq V\leq V_{a\;max}$ are related.

\begin{equation}\label{eq3}
T=\frac{a}{V_{a\;min}} -\frac{a}{V_{a\;max}}.
\end{equation}

Distance $a$ and the maximum velocity $V_{a\;max}$ captured by the lens determine the minimum travel time from the source to the lens
\begin{equation}\label{eq4}
t_{a min}=t_1=\frac{a}{V_{a\;max}}.
\end{equation}

The minimum velocity captured by the lens and the maximum time of flight of distance $a$ are determined as
\begin{equation}\label{eq5}
V_{a\;min}=\frac{a V_{a\;max}}{T V_{a\;max}+a}\;\;\;t_2=\frac{a}{V_{a\;min}}.
\end{equation}

Velocities  $V_{a\;max}$ and $V_{a\;min}$ determine the magnitude of the neutron flux transformed by the lens. The range of captured velocities can be called “velocity aperture of the lens”. The value of the maximum neutron velocity after the lens  $V_{b\;max}$ is limited by the properties of the trap that accumulates neutrons. The minimum flight time of the second part of the neutron guide $t_{b\;min}=b/ V_{b\;max}$ is also determined by setting this velocity.
The neutrons that have spent most of the time in the first part of the path should obviously have the highest velocities after the lens. Therefore, the total flight time $t_0$ is
\begin{equation}\label{eq6}
t_0=t_{a\;max}+t_{b\;min}=\frac{a}{V_{a\;min}} +\frac{b}{V_{b\;max}}.
\end{equation}

The maximum energy transfer is defined by velocities $V_{a\;max}$ and $V_{b\;min}$
\begin{equation}\label{eq7}
\Delta E=\frac{m}{2} \left[V_{a\;max}^2-V_{b\;min}^2\right].
\end{equation}

The most important question is how to change the neutron energy according to a given time law~(\ref{eq2}). As an attractive, but seemingly somewhat extravagant possibility, in~\cite{AIFrank1996, AIFrank2000}, it was proposed to turn to non-stationary quantum phenomena. Among the latter, phase modulation of a neutron wave by a phase diffraction grating moving across the direction of its propagation and the resonant spin-flip of the neutron in a magnetic field were considered. Both of these proposals were quite ambitious at that time, since the phenomenon of non-stationary neutron diffraction on a moving grating was only a subject of theoretical consideration~\cite{AIFrankPLA1994}, and the change in the neutron velocity during a resonant spin-flip, which was previously observed at the limit of experimental possibilities~\cite{Alefeld1981}, was more convincingly demonstrated in~\cite{Weinfurter1988} only after the publication of work~\cite{AIFrank1996}.

Later on, the non-stationary diffraction of UCNs by a moving grating was observed in the experiment described in~\cite{AIFrankPLA2003}, and after some time, using aperiodic moving grating, the effect of focusing in time was demonstrated in experiments~\cite{AIFrankJETPL2003, Balashov2004}. The possibility of time focusing based on the resonant spin-flip of the neutron has also been experimentally confirmed~\cite{Arimoto2012, Imajo2021}.

\subsection{Moderation of neutrons by a lens}

Let us now explain the idea of F. L. Shapiro~\cite{Shapiro1976} about the accumulation of neutrons from a pulsed source. Supposedly, we have an ultracold neutron trap with an entrance hole, the area of which is $s$, where a periodic pulsed flux with pulse duration $\tau$ enters. Therefore, the number of neutrons entering the trap per pulse is
\begin{equation}\label{eq8}
N_{in}=F_s s \tau,
\end{equation}
where $F_s$ is the neutron flux averaged over the pulse duration. Let us assume that the entrance window of the trap is open only during the pulse, and for the rest of the time it is blocked by some ideal shutter. If during the time between the pulses the number of neutrons absorbed in the trap walls is relatively small, the neutron flux in the trap will grow from pulse to pulse until it reaches some equilibrium value $\Phi$. Neglecting the probability of neutron decay during the time of equilibrium establishment, we assume that the channels of UCN escape from the trap are only absorption in its walls and escape through the entrance hole during the time when it is open. Equating the number of neutrons entering the trap in one pulse~(\ref{eq8}) to the number of neutrons leaving it during the time of one cycle $T$, we obtain:
\begin{equation}\label{eq9}
F_s S \tau=\Phi(s \tau+\Sigma T \mu),
\end{equation}
where $\Sigma$ is the area of the trap surface, $\mu$ is the coefficient of UCN absorption at a collision with the trap wall, and $T$ is the period of pulses. Consequently, for the equilibrium flux density we have
\begin{equation}\label{eq10}
\Phi=F_s \frac{1}{1+\left[\Sigma T \mu/s \tau \right]},
\end{equation}

From equation~\ref{eq10}, it is easy to obtain expressions for the ratio of the flux density in the trap to the average flux at the trap entrance $<F>=F_s \tau/T$, that is, the gain factor.
\begin{equation}\label{eq11}
G=\frac{\Phi}{<F>},\;\;\; G=\frac{s T}{s \tau +\Sigma \mu T}.
\end{equation}

Equation~(\ref{eq11}) is slightly different from formula~(8) of work~\cite{Shapiro1976}, which is due to two circumstances. Firstly,~(\ref{eq11}) does not take into account the possibility of neutron outflow from the trap to the user. Secondly,~\cite{Shapiro1976} compares the neutron flux density in the trap in the case of pulsed and stationary filling methods, rather than the ratio of fluxes in the source and in the trap.

Absorption coefficient $\mu$, when a neutron reflects from the wall, is determined by the ratio $\eta=W/E_b$ of the imaginary and real parts of the effective potential of the neutron and trap substance interaction.

\begin{equation}\label{eq12}
U=E_b-i W,\;\;\;E_b=\frac{2 \pi \hbar^2}{m}\rho b,\;\;\;W=\frac{\hbar}{2}\rho \sigma_{loss} v.
\end{equation}

Here $b$ is the neutron scattering length of the trap material, $\rho$ is the nuclear density, $\sigma_{loss}$ is the cross-section of all processes in the substance of the trap leading to the disappearance of the UCNs from it, and v is the neutron velocity. The real part of the potential $E_b$ is usually called the boundary energy of the substance. For an isotropic flux, the angle-averaged absorption coefficient $\mu$ is~\cite{Shapiro1976, Ignatovich1990}
\begin{equation}\label{eq13}
\mu=\frac{2\eta}{y^2}\left(\arcsin y -y\sqrt{1-y^2}\right).
\end{equation}
where $y=\sqrt{E/E_b}$, and $E$ is the neutron energy. If energy $E$ is not too close to the boundary energy, values $\mu$ and $\eta$ differ slightly over a wide range of changes in the parameter $y$. For a number of good materials, $\eta$ can be $\eta=(3\div5)\times10^{-5}$~\cite{Nesvizhevskii1992, Arzumanov2003, Atchison2007} and the parameter $\Sigma \mu T/s \tau$ in the denominator of equation~(\ref{eq10}) can be of the order of unity even for a trap of quite significant volume. In this case, gain factor   turns out to be of the order of   As an example, we note that for the IBR-2 reactor, the ratio is $T/\tau\approx500$.

\subsection{Neutron pulse time transformation — time magnification}

Since gain factor  significantly depends on the duration of the open state of the trap gate, the question of the pulse duration formed by the time lens is very important. In~\cite{AIFrank2000}, it was assumed that at a relatively small energy transfer by the lens $|\Delta E|\ll E$, the formula of a thin lens can be used
\begin{equation}\label{eq14}
M=\tau_{fin}/\tau_{in},\;\;\;M=b/a,
\end{equation}
where $\tau_{in}$ and $\tau_{fin}$ are the duration of the initial time pulse and its "image" formed by the lens, respectively. However, this is true only in the case when the focusing conditions are symmetric (Fig.~\ref{subfig:Fig1a}), that is, when a certain amount of neutrons accelerates, and the same amount decelerates. In the case of the moderating lens, it is not so. The question of the relation between the pulse duration of a neutron flux generated by a neutron source and its time image was considered in more detail in~\cite{AIFrank2021}.

It was shown that neutrons emitted not exactly at the moment of time $t=0$, but at a moment  close to it, arrive at the observation point not at the calculated time, but at the moment
\begin{equation}\label{eq15}
t=t_0+\delta \frac{b}{a}\left(\frac{V_a}{V_b}\right)^3,
\end{equation}

Assuming the smallness of the lens action period (time aperture) $t_1 - t_2 \ll t_L$, which is similar to the paraxial approximation in optics, it is easy to obtain the following ratio for time magnification
\begin{equation}\label{eq16}
M=\frac{b}{a}\left(\frac{V_a^0}{V_b^0}\right)^3,
\end{equation}
where $V_a^0$ and $V_b^0$ are the averaged values of the velocities in sections a and b, respectively.

Thus, the duration of the image formed by the time lens depends not only on the geometric factor  , but also essentially on the ratio of the initial and final velocities, and the shape of the time pulse can also be determined by the velocity spectra before and after the lens. A considerable time magnification value for the moderating lens significantly limits the value of the gain factor~(\ref{eq11}) due to the accumulation mode of the trap.

\subsection{Non-stationary neutron diffraction as a physical basis for energy transformation by a time lens}

The effect of the local time lens is based on the possibility of a strictly defined and time-variable energy transfer to the neutron. Two methods of energy transfer based on the use of quantum non-stationary phenomena were mentioned above. The first is the spin-flip of the neutron by a spin-flipper, the mandatory component of which is an alternating magnetic field with an angular frequency $\Omega$, the other one is the flux modulation by a device that provides an amplitude or phase modulation of the flux with a frequency of $f=\Omega/2\pi$. In both cases, energy equal to or a multiple of the energy quantum $\Delta E=\hbar \Omega$ is transferred to the neutron. One would assume that a flipper is more efficient than a modulator. In fact, all neutrons whose spin has been flipped under the action of an alternating field with frequency $\Omega(t)$ in the presence of a relatively slowly changing magnetic field $B(t)$ will change their energy by $\Delta E(t)=\mu B(t)=\hbar \Omega(t)$.

In contrast to the case above, at periodic modulation of the flux, the resulting state has the form of a superposition of a large number of waves with different amplitudes $A_n<1$ and discrete frequency values $\omega_n=\omega_0+n\Omega$, where $n$ is integers, regardless of whether such modulation is carried out using a fast periodic chopper~\cite{Nosov1991,AIFrankPAN1994}, or a periodic structure moving across the beam~\cite{AIFrankPLA1994}. From the complete set of waves at any given time, only one has the energy that meets the necessary condition for the time focusing, and its amplitude depends both on the parameters of the modulator, and, generally speaking, on the time-varying velocity of the incident neutrons. It would seem that under these conditions, the task of choosing between two approaches to controlling the neutron energy should be solved in favor of the electromagnetic method.

However, from a practical point of view, the following two factors are important: the maximum possible value of the transferred energy and the simplicity of a quick change in the required frequency. The latter should change from its maximum to the minimum value during the period of action of the source and lens, and at the same time almost instantly reset to the initial value at the end of each cycle. It turns out that both of these conditions are currently better met by a device based on the phenomenon of non-stationary diffraction on a moving structure.

It is important that the phenomenon of non-stationary neutron diffraction by a moving grating is well-studied theoretically and experimentally. There are several theoretical approaches to solving the problem of neutron diffraction by a moving grating. Here, we present a simplified version of the solution to this problem following works\cite{AIFrankPLA1994, AIFrankPLA2003,AIFrank2005}.

Let a plane neutron wave fall on a thin grating, the groves of which are oriented along the $y$-axis
\begin{equation}\label{eq17}
\Psi_0(x,z,t)=\exp[i\left(k_{0x}x+k_{0z}z-\omega_0 t\right)]
\end{equation}
where $k_{0x}=mv_{0x}/\hbar$ and $k_{0z}=mv_{0z}/\hbar$ are the tangential and normal components of the velocity, respectively, $\hbar$ is the Planck constant, $\omega_0=\hbar k_0^2/2m$ and $k_0=\left(k_{0x}^2+k_{0z}^2\right)^{1/2}$ are the frequency and wave number.

Let us assume that the grating moves with the velocity $V_g$ in the positive direction of the x-axis. Solving the problem of diffraction in a moving system, in which the grating is at rest, we can find the projections of the wave vectors of all diffraction orders and their amplitudes $\alpha_n$
\begin{equation}\label{eq18}
\alpha_n=\frac{1}{d}\int_{0}^{d} H(x)\exp \left(-ing_0 x\right)dx,
\end{equation}
where $g_0=2\pi/d$, $d$ is the grating spatial period, and $H(x)=H(x+d)$ is the periodic transmission function of the grating. The wave function of the diffracted neutrons is found by transforming the resulting solution back to the laboratory coordinate system. It has the form
\begin{equation}\label{eq19}
\Psi_0(x,z,t)=\sum_{n=-\infty}^{\infty}A_n\exp[i\left(k_{nx}x+k_{nz}z-\omega_n t\right)],
\end{equation}
where
\begin{equation*}
\begin{split}
&k_{nx}=k_{0x}+g_n,\;\;k_{nz}=\left[k_{0z}^2+2\left(k_V-k_{0x}\right)g_n-g_n^2\right]^{1/2},\\&k_V=mV_g/\hbar,
\end{split}
\end{equation*}
$g_n=n g_0$ is the value of the reciprocal lattice vector, $n=0,\pm 1,\pm 2, ...$ is integers. The frequencies of the diffracted waves $\omega_n=\omega_0+n\Omega$ are characterized by spectral splitting
\begin{equation}\label{eq20}
\Omega=2\pi f,\;\;\;\text{where}\;\;f=V_g/d.
\end{equation}

The amplitudes of the diffracted waves in the laboratory system are determined from the flux conservation condition
\begin{equation}\label{eq21}
A_n=\alpha_n\left[k_0^2/\left(k_0^2+2k_V g_n\right)\right]^{1/4}.
\end{equation}

In the early works~\cite{AIFrankPLA2003, AIFrank2005}, a very satisfactory agreement with the predictions of such a theory was obtained. As suggested in~\cite{AIFrankPLA1994}, the experiments were performed with a grating having a rectangular profile, for which the phase of the transmitted wave changed abruptly by $\pi$ at every half-period.
\begin{equation}\label{eq22}
\Delta\phi=k_{0z}\left(1-n_{gr}\right)h=\pi.
\end{equation}
Here $h$ is the height of the grating tooth and $n_{gr}$ is the refractive index of neutrons in the material of the grating. For such a phase $\pi$-grating, the amplitudes of even orders, including zero, are equal to zero, the amplitudes of odd orders are equal to   and decrease with an increasing order number n. However, this is true only when neutrons fall on it normally, which is possible for a stationary grating, since in the coordinate system of a moving grating, neutrons fall on it at a certain angle. Therefore, in fact, the phase change has a trapezoidal form due to the different path lengths in the substance near the edges of the grating rectangular teeth.
\begin{figure}[hb]
\centering
        \includegraphics[width=\columnwidth]{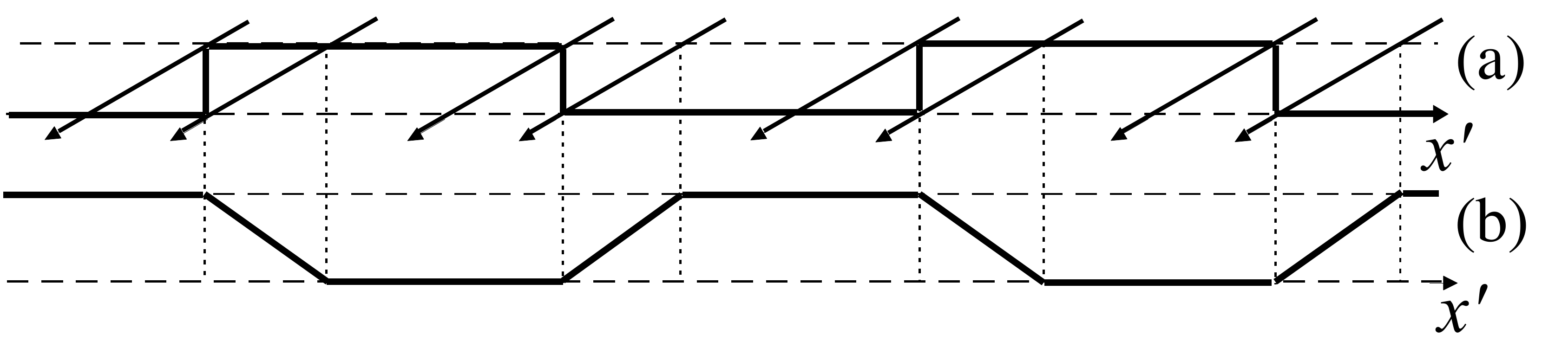}
\caption{(a) Grating profile, (b) Phase profile at oblique incidence on the grating.}
\label{fig:Fig2}
\end{figure}

This circumstance was taken into account in~\cite{AIFrank2004}, where the trapezoidal dependence of the phase on the coordinate was used to calculate the amplitudes of the diffraction orders in accordance with~(\ref{eq14}) (see Fig.~\ref{fig:Fig2}). The profile of the phase dependence was characterized by a geometric parameter
\begin{equation}\label{eq23}
C=\left(2h/d\right)\left(V_g-v_{0x}\right)/v_{0z},
\end{equation}
which increases with an increasing grating velocity, profile depth, and decreasing grating period. At $C\ll 1$ the phase profile is close to rectangular, and at $C=1$ it takes a triangular shape.

In a more rigorous approach, it is necessary to take into account that waves of different diffraction orders propagating in a grating material of finite thickness can mutually influence each other. To take into account this circumstance, a dynamical approach to diffraction formulated in~\cite{Bushuev2016} is necessary. In this work, a very important result was obtained, which consists in the fact that for a given grating velocity and neutron energy, the ratio between the intensities of orders depends significantly on the depth of profile $h$. Therefore, abandoning the fulfillment of relation~(\ref{eq22}), it is possible to set the depth of profile $h$, and, together with it, parameter $C$, proceeding from the specific values of the grating velocity and the value of the transmitted energy. A number of predictions of work~\cite{Bushuev2016} were confirmed in experiments~\cite{Kulin2016, Kulin2019}.   

We should also note that although for the amplitudes of diffraction orders the dynamical theory predicts slightly different values than the modified kinematic approach, both theories give qualitatively similar results~\cite{Bushuev2016}. 

\section{UCN source for a periodic pulsed reactor}

\begin{figure*}[h]
\centering
        \includegraphics[width=0.75\linewidth]{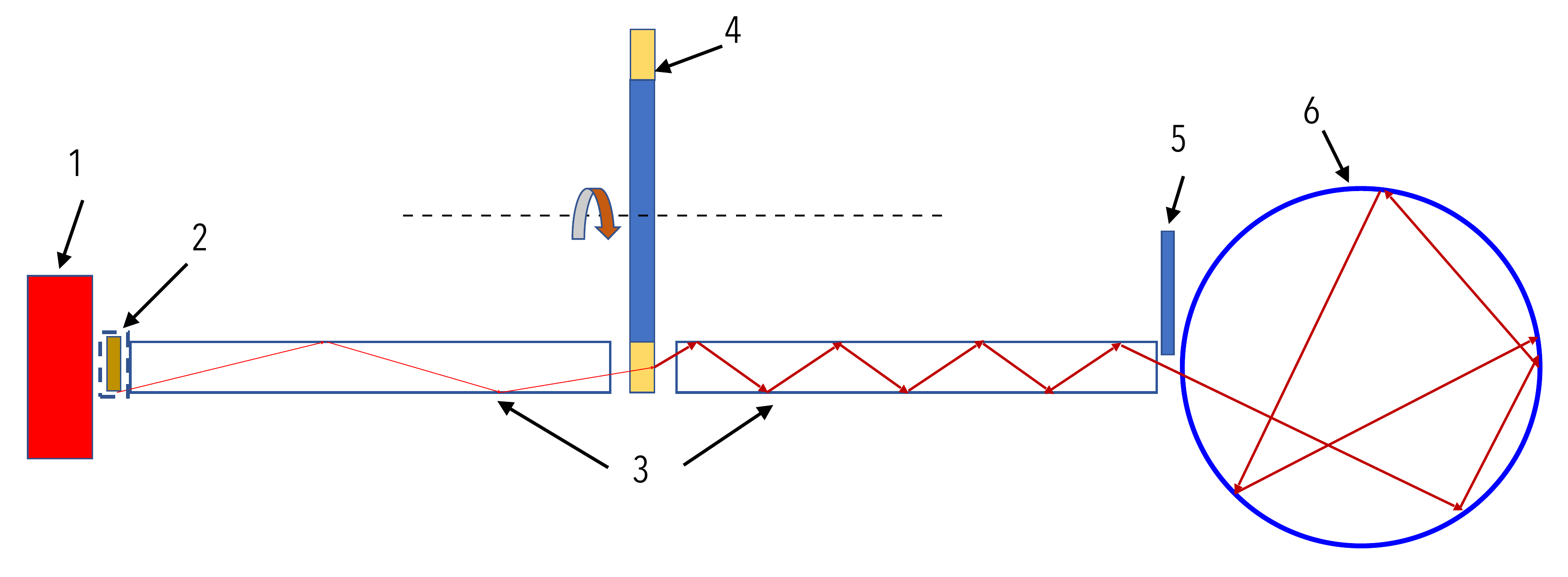}
\caption{Schematic diagram of the source. 1 - main moderator, 2 – converter moderator, 3 - neutron guide, 4 - time lens with diffraction gratings, 5 - fast valve, 6 - storage trap.}
\label{fig:Fig3}
\end{figure*}

A schematic diagram of a UCN source based on the time-focusing principle is shown in Fig.~\ref{fig:Fig3}. The main moderator 1 is a source of a pulsed cold neutron flux. The thin converter moderator 2 serves as a source of UCN pulses. Its thickness is bounded by the product of the minimum velocity captured by the lens and the required duration of the UCN pulse flux. It can be separated from the volume of the mirror neutron guide 3 by a membrane transparent to UCNs. The need for background suppression probably precludes the use of a rectilinear neutron guide, but for now we will ignore this circumstance.

The time lens 4 is a set of diffraction gratings located at the periphery of a rotating disk. The rotation period of the lens disk is equal to or a multiple of the repetition period of the pulsed neutron source. The rotation phase is synchronized with the source. The grating parameters are optimized to provide the focusing condition.

Let us temporarily ignore the practically important question of the arrangement of the pulse valve at the entrance to the trap and restrict ourselves to analyzing the transfer and transformation properties of the combination of neutron guides and the lens, the principle of which is based on the non-stationary diffraction of neutrons by a moving grating. We present the results of some calculations that allow us to estimate the possible parameters of such a source. To be more specific, we have proceeded from the parameters typical of the IBR-2 reactor~\cite{Ananiev1977, Dragunov2012}.

\subsection{UCN flux in the converter}

The pulsed flux of thermal and cold neutrons is formed by the main moderator 1 of the source. The energy distribution of neutrons is assumed to be Maxwellian with an effective temperature $T_n$.  Consequently, the average UCN flux\footnote{Hereinafter, as it is conventional in neutron physics, we will use the terms "flux" and "flux density" in the same sense, in both cases meaning the flux density with the dimension cm$^{-2}$s$^{-1}$.} with energies less than a certain boundary energy $E_b$ is~\cite{Dragunov2012}
\begin{equation}\label{eq24}
<F>=\frac{<\Phi_0>}{8}\left(\frac{E_b}{T_n}\right)^2,
\end{equation}
where $<\Phi_0>$ is the total average flux in the moderator. In certain cases, the UCN flux in the converter may exceed the flux in the moderator, which is taken into account by introducing a certain gain factor (as an example, see~\cite{Golikov1973}).

The average thermal neutron flux in the IBR-2M reactor is  $\Phi_0=2\times10^{12}$cm$^{-2}$s$^{-1}$. Taking a value of 400 K as the temperature of the Maxwell neutron spectrum and $Eb=190$ neV for the boundary energy, for the density of the UCN flux we obtain the value  $\Phi=8$cm$^{-2}$s$^{-1}$. The choice of the specified value $E_b$ is determined by setting the maximum neutron velocity at the entrance to the trap (see below). Probably, it is possible to use a converter in which the UCN flux density exceeds the flux density in the external moderator by one order of magnitude. Such a converter is, for example, ice at low temperature~\cite{Ananiev1989}. This results in estimating $\Phi=80$cm$^{-2}$s$^{-1}$.

\subsection{Lens parameters, transport time, and neutron guide transmission}

As it was assumed in the calculations, the maximum longitudinal velocity after the lens should not exceed 3 m/s, so that the total velocity at the entrance to the trap was less than or of the same order of magnitude as the boundary velocity for beryllium $V_b^{Be}=6.9$ m/s, which we chose as the trap material. The total neutron transfer time depends on the position of the lens and the maximum velocity captured by it. With a neutron guide length $L=10$ m, this is 2-3 seconds. 
The calculations were carried out by the Monte Carlo method. It was assumed that the flux in the moderator is isotropic, the neutron velocities are limited to a certain value $V_{max}$, and the probability of finding a neutron in the velocity range from $V$ to $V+dV$ is proportional to $V^3$. The choice of the quantity $V_{max}$ is rather arbitrary, since the result was normalized to the standard phase volume determined by the boundary velocity of the trap (see below). It is only necessary to fulfill the condition $V_{max}>V_{a\;max}$, where $V_{a\;max}$ is the maximum longitudinal velocity captured by the lens. In the calculation it was assumed that $V_{max}=\sqrt{2}V_{a\;max}$.

Velocity $V_{a\;max}$, together with the distance to the lens, determines the time of neutron transport from the moderator to the trap. Since the length of the neutron guide is much larger than its transverse diameter, practically all neutrons for which the velocity component $v_{\perp}$ normal to the axis of the neutron guide is greater than its boundary velocity are not captured by the neutron guide and are lost. Neutrons with a lower transverse velocity experience multiple collisions with the walls, with some probability being absorbed in them, and propagate in the direction of the lens and the trap. As the probability of absorption of a neutron with a velocity normal to the surface of the substance, not too close to the boundary, is~\cite{Ignatovich1990}
\begin{equation}\label{eq25}
\mu=2\eta\frac{v_{\perp}}{\sqrt{v_{bg}^2-v_{\perp}^2}},
\end{equation}
the probability of successful neutron transport is determined by the expression
\begin{equation}\label{eq26}
\vartheta=\left(1-2\eta_{ng}\frac{v_{\perp}}{\sqrt{v_{bg}^2-v_{\perp}^2}}\right)^n,
\end{equation}
where $n$ is the number of collisions with the walls of the neutron guide and $\eta_{ng}$ is the parameter of neutron absorption by the neutron guide substance. Since, under the conditions of time focusing, the lens provides isochronous neutron transport, the number of collisions of neutrons, which have reached the trap, with the walls is 
\begin{equation}\label{eq27}
n=\frac{v_{\perp}}{D}t_0,
\end{equation}
where $D$ is the transverse size of the neutron guide and $t_0$ is the total transport time through both sections of the neutron guide. According to the calculations, the diameter of the neutron guide is $D=8$ cm, the boundary velocity is  $v_{bg}=6.5$ m/s, and the absorption parameter is $\eta_{ng}=10^{-4}$. These values correspond to the parameters of the non-magnetic alloy NiMo, which is widely used in the UCN physics. 

The calculations have shown that under the assumption made about the specularity of the neutron guides, the losses of neutrons absorbed in the walls do not exceed a few percent, and the transport efficiency is almost completely determined by the fraction of the phase volume captured by the system. The latter depends on the choice of value $V_{max}$, boundary velocity $v_{bg}$, and the velocity aperture of the lens.

\subsection{Diffraction efficiency of a time lens and transfer properties of the entire system}
\begin{figure*}[h]
\centering
\begin{minipage}{\columnwidth}
        \includegraphics[width=0.95\textwidth]{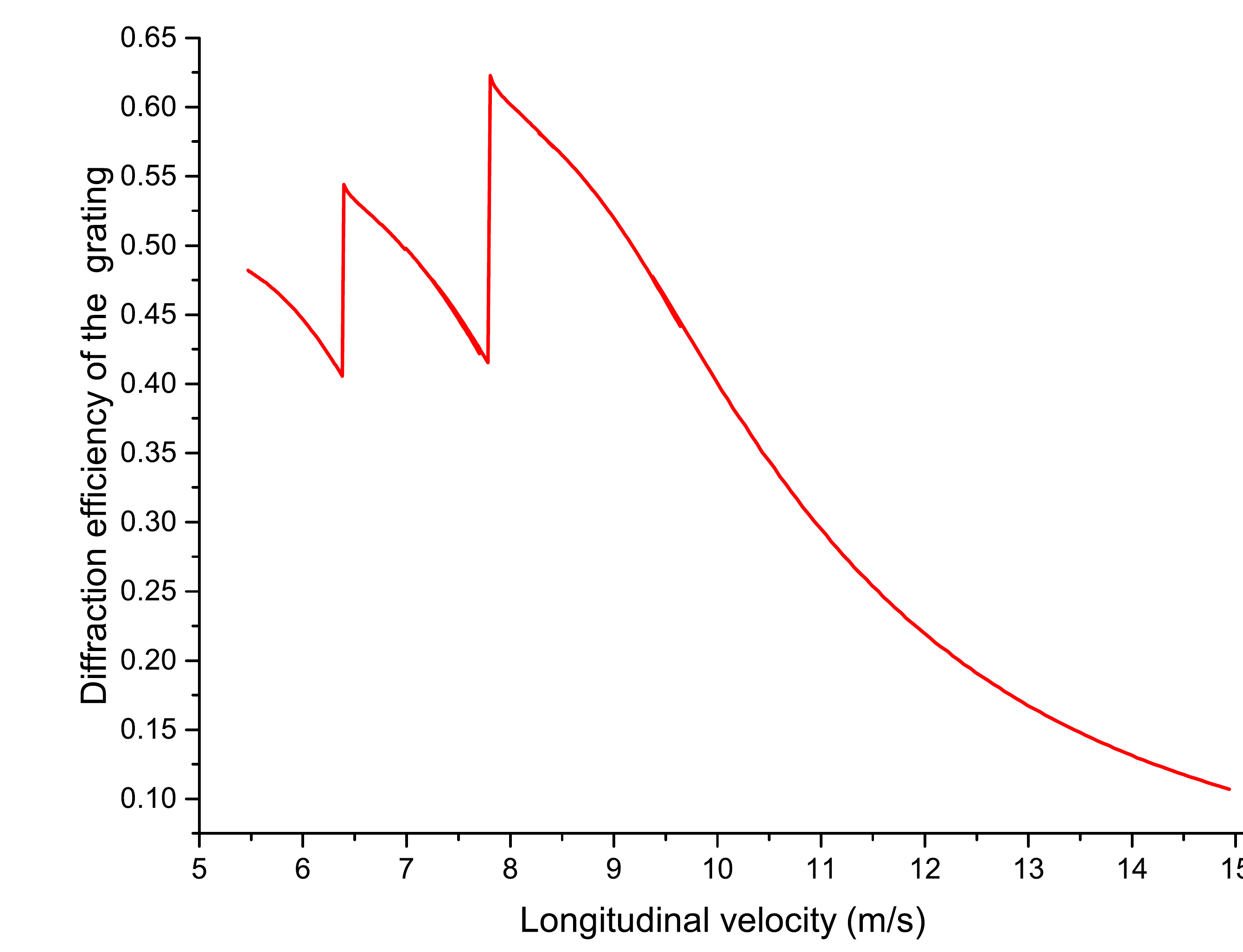}
         \caption{Diffraction efficiency of the lens as a function of the longitudinal neutron velocity.}  \label{fig:Fig4}
    \end{minipage}
   \hfill
    \begin{minipage}{\columnwidth}
        \includegraphics[width=0.95\textwidth]{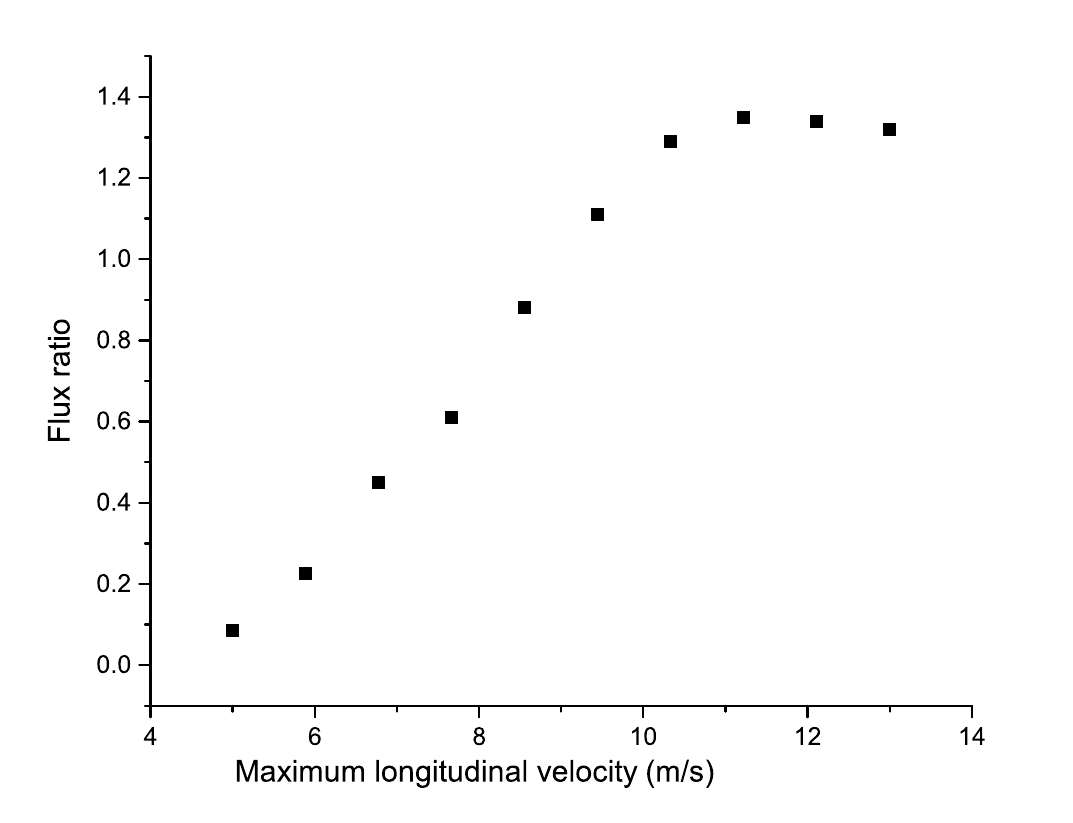}
          \caption{Transport efficiency of a channel with a lens as a function of the maximum longitudinal neutron velocity (see the text).}  \label{fig:Fig5}
    \end{minipage}
\end{figure*}
The calculations of the diffraction efficiency of the lens were carried out in a modified kinematic approximation. The incidence of neutrons on the grating in the laboratory system was assumed to be normal. The depth of the grating profile and the distance between the slits varied continuously depending on the current value of the velocity of the incident neutrons $V_1$. Neglecting the tangential velocity with respect to the grating, equation~(\ref{eq23}) takes the form
\begin{equation}\label{eq28}
C=\frac{2h}{V_1}\frac{V_g}{d}=\frac{2h}{V_1}f,
\end{equation}
where $f=V_g/d$ is the modulation frequency specified by the current value of the required energy transfer~(\ref{eq2}) and relation $\Delta E(t)=2\pi f(t) h$. It was assumed that the time dependence of the frequency is provided by a continuous change in the slowly varying parameter $d$, at a constant grating velocity. Knowing the value of $f$ at each instant, from equation~(\ref{eq28}) it is possible to obtain the maximum height of the grating tooth, for which the limiting relation   $C=1$ is satisfied. In this case, the phase modulation function had a trapezoidal form~\cite{AIFrank2004}. In contrast to the case shown in Fig.~\ref{fig:Fig2}, the phase was changing from zero to a certain value $\phi$, which was determined by the neutron velocity, the height of the tooth, and the grating material with the boundary velocity $V_{bgrat}$
\begin{equation}\label{eq29}
\phi=k(1-n)h=\frac{2\pi}{\lambda}\left[1-\sqrt{1-\frac{V_{bgrat}^2}{V_1^2}}\right]h.
\end{equation}

As can be shown, the amplitude of the minus first order wave is described in this case by the formula
\begin{equation}\label{eq30}
\alpha_{-1}=\frac{i\phi}{2\pi^2}\frac{\left(e^{i\phi}e^{i\pi C}-1\right)\left(\frac{\phi}{\pi}-C\right)-\left(e^{i\pi C}-e^{i\phi}\right)\left(\frac{\phi}{\pi}+C\right)}{\frac{\phi^2}{\pi^2}-C^2}.
\end{equation}

The diffraction efficiency, that is, the intensity of the minus first order wave $|A_{-1}|^2$was calculated using equations~(\ref{eq30}),~(\ref{eq21}) for a large set of values $0<h<h_{max}$. Based on their results, the value of $h$ was chosen to provide the maximum diffraction efficiency. For different ranges of neutron velocity, one of four values $V_{bgrat}$ was chosen from 3.2 m/s (Si) to 7.8 m/s ($^{58}$NiMo alloy).

The results of calculating the diffraction efficiency of the lens are illustrated in Fig.~\ref{fig:Fig4}. The efficiency jumps seen in this figure are due to transitions to a new grating material with a different value $V_{bgrat}$. Note that in accordance with equation~(\ref{eq21}) the intensity of diffraction orders corresponding to neutron moderation can be greater than for a grating at rest, which does not change the velocity spectrum. This circumstance is taken into account in the results presented in Fig.~\ref{fig:Fig4}.

\begin{figure*}[h]
\centering
\begin{minipage}{\columnwidth}
        \includegraphics[width=0.95\textwidth]{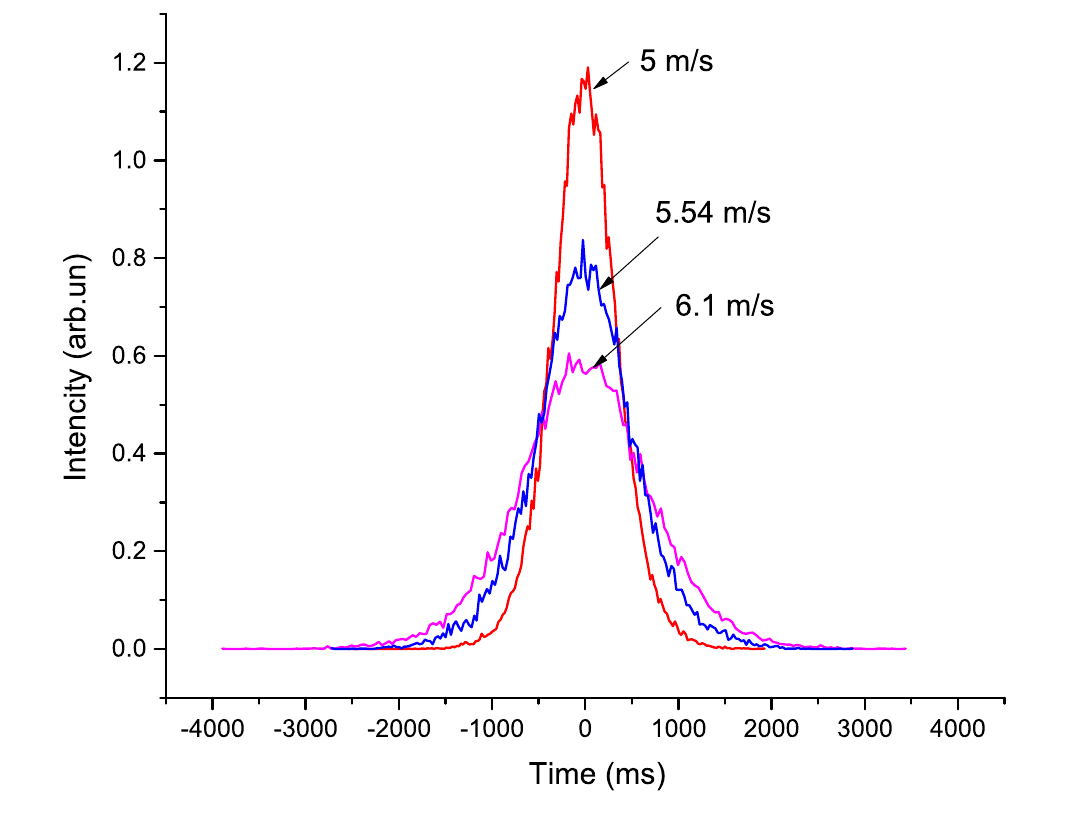}
         \caption{Results of calculating the shape of the pulse formed by the lens for three values of the maximum longitudinal velocity.}  \label{fig:Fig6}
    \end{minipage}
   \hfill
    \begin{minipage}{\columnwidth}
        \includegraphics[width=0.95\textwidth]{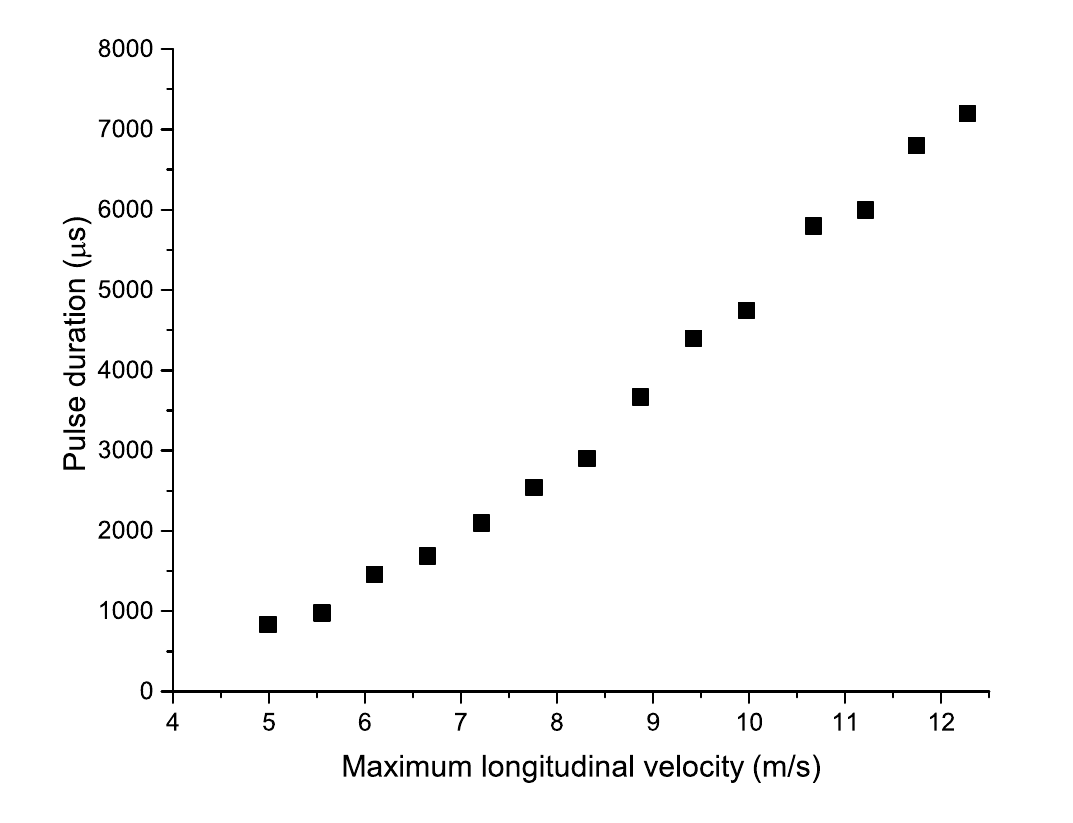}
          \caption{Width of the pulse formed by the lens, depending on the maximum longitudinal velocity.}  \label{fig:Fig7}
    \end{minipage}
\end{figure*}
\begin{figure*}[h]
\centering
\begin{minipage}{\columnwidth}
        \includegraphics[width=0.95\textwidth]{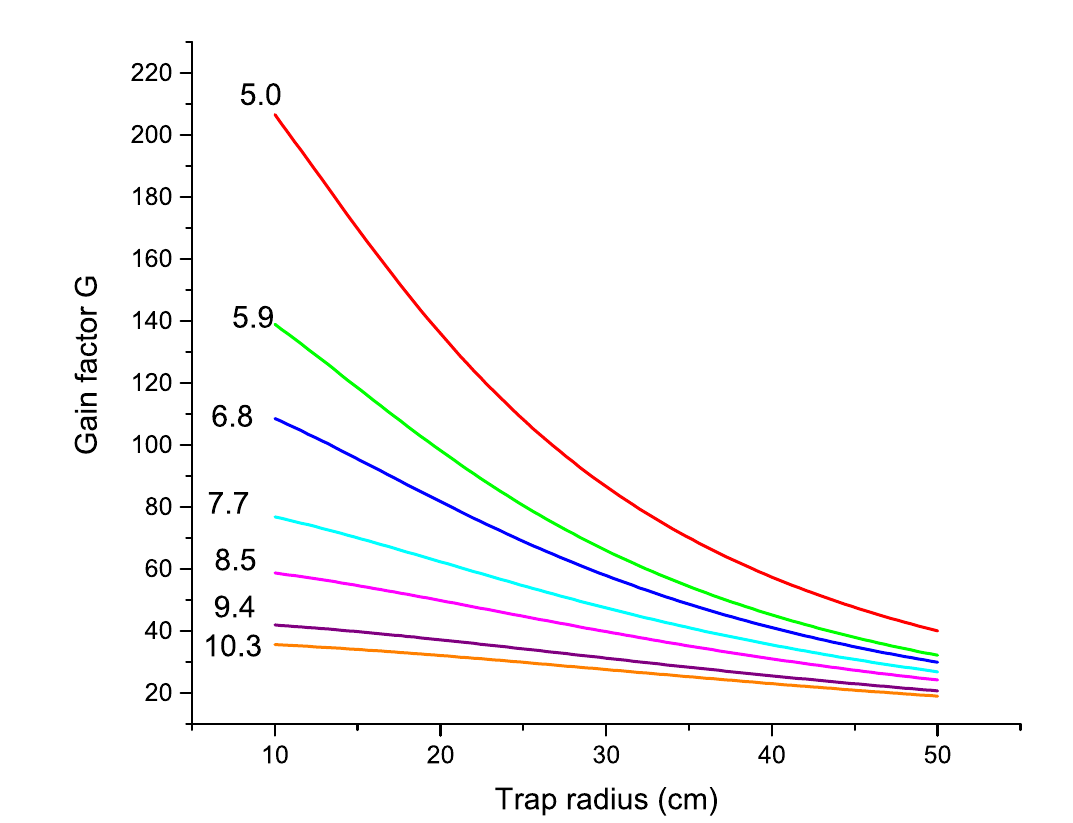}
         \caption{Results of calculating the shape of the pulse formed by the lens for three values of the maximum longitudinal velocity.}  \label{fig:Fig8}
    \end{minipage}
   \hfill
    \begin{minipage}{\columnwidth}
        \includegraphics[width=0.95\textwidth]{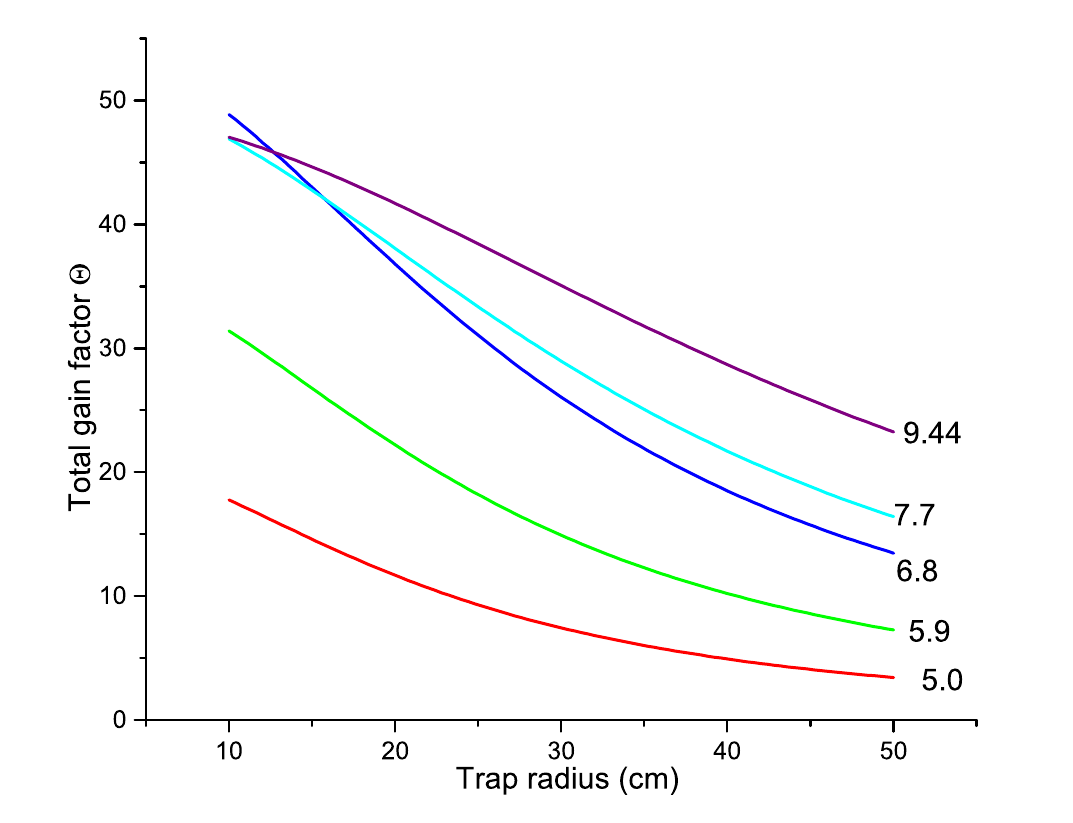}
          \caption{Width of the pulse formed by the lens, depending on the maximum longitudinal velocity.}  \label{fig:Fig9}
    \end{minipage}
\end{figure*}

The diffraction efficiency of the lens and the efficiency of capture and transport of neutrons determine the value of the efficiency of the entire system $T$. When determining this value, it was taken into account that the calculation of the probability of transport was carried out for the phase volume of velocities occupied by a sphere with a radius $V_{max}=\sqrt{2}V_{a\;max}$, and the neutron flux density in such a phase volume exceeds the UCN flux density with velocities from zero to $V_{bn}$ in the ratio $K=\left(V_{a\;max}/V_{bn}\right)^4$, where $V_{bn}=\sqrt{2E_b/m}$ (see eq.~(\ref{eq26})). Thus, the transport efficiency of a system consisting of two neutron guides and a lens is understood as the ratio of the flux at the exit from the neutron guide, that is, at the entrance to the trap, to the flux of neutrons with velocities less than $V_{bn}=6$ m/s in the converter. The dependence of efficiency $T$ determined in this way on the maximum longitudinal velocity is illustrated in Fig.~\ref{fig:Fig5}. The minimum velocity captured by the lens was calculated using equation~(\ref{eq5}).

\subsection{Pulse duration, accumulation effect, and intensity of the UCN source}

Depending on the selected value of the maximum longitudinal velocity, the ratio of the velocities before and after the lens will change. This means that, in accordance with~(\ref{eq15}), each value of the longitudinal velocity corresponds to its own value of the deviation in the time of arrival $t$ at the trap from the estimated time $t_0$. Therefore, the shape of the resulting pulse was calculated taking into account the distribution of longitudinal velocities at the exit of the neutron guide. The initial pulse shape was specified as a Gaussian with a half-width of 350 $\mu$s.
The results of such calculations are illustrated in Figs.~\ref{fig:Fig6} and~\ref{fig:Fig7}. For the pulse duration in Fig.~\ref{fig:Fig7}, the width at half maximum was taken. The pulse duration, together with the specified trap parameters, determines the value of the gain from the pulse accumulation mode. Therefore, the latter now depends on the maximum longitudinal speed.

Figure~\ref{fig:Fig8} shows the dependence of gain factor G, (see eq.~(\ref{eq11})), due to the pulsed nature of the trap filling, depending on its radius for several values of the maximum longitudinal velocity. The calculation takes into account the increase in the duration of the pulse generated by the lens. For the absorption parameter of the trap material, a value  was taken, which corresponds to the experimental data for a beryllium trap at low temperatures~\cite{Nesvizhevskii1992, Ageron1985}.

However, the total efficiency of the source, that is, the ratio of the UCN flux density in the trap to the UCN flux density in the converter, is characterized not by factor $G$, but by its product by the value of the transport efficiency of channel $T$, which can be greater than unity for a lens that slows down neutrons. The dependence of the total gain factor $\Theta=GT$ on the trap radius is shown in Fig.~\ref{fig:Fig9}. As can be seen from this figure, the gain factor increases with an increase in the longitudinal neutron velocity, which is caused by a certain compression of the flux by the moderating lens. However, with a further increase in the speed, this growth stops, which is due to the drop in the diffraction efficiency of the lens, which is clearly seen in Fig.~\ref{fig:Fig4}.  Multiplying the gain factor by the value of the average flux in the moderator, we obtain an estimate of the UCN flux in the trap, which, in essence, is the main result of the calculation.

\begin{figure}[h]
\centering
        \includegraphics[width=\columnwidth]{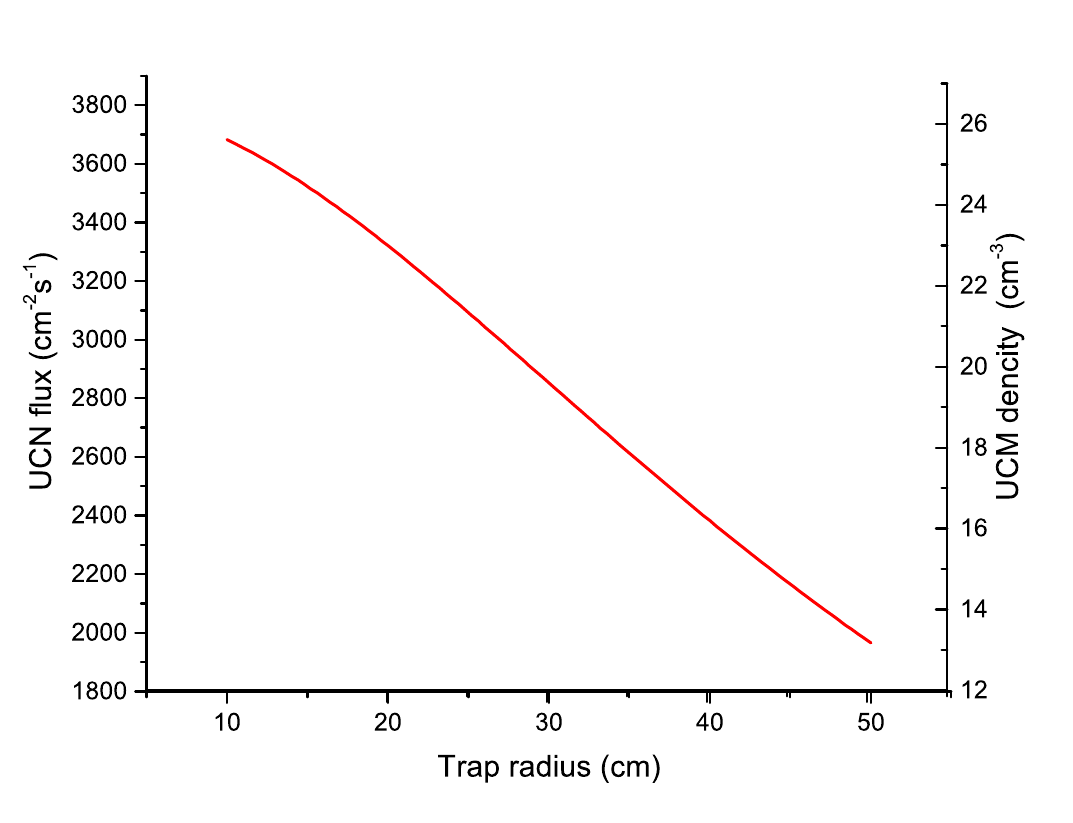}
         \caption{Flux density and volume density of UCNs in a trap depending on its radius.}  \label{fig:Fig10}
\end{figure}

\section{Discussion of the results}

The graph in Fig.~\ref{fig:Fig10} demonstrates that the principle of the focusing in time accompanied by neutron moderation makes it possible to create a sufficiently intense UCN source in a periodic pulsed reactor even with a moderate value of the average neutron flux.
The above-given results, apparently, demonstrate the potential of the discussed concept, but at the same time they are rather evaluative. Let us summarize the main parameters of the calculation and the approximations made.
\begin{enumerate}
\item The average flux of thermal neutrons is $\Phi_0=2\times10^{12}$ n/ cm$^2$s.
\item The effective temperature of the Maxwellian spectrum of neutrons is T = 400 K.
\item The converter is characterized by a gain factor of 10.
\item The neutron guide is absolutely specular.
\item The calculation of the diffraction efficiency of gratings is based on a modified kinematic approximation.
\item A pulse valve at the entrance to the trap is ideal.
\item The boundary velocity and absorption parameter of the trap material are 6 m/s, and $\eta=3\times10^{-5}$, respectively.
\end{enumerate}

These approximations lead to opposing factors distorting the result. Thus, items 4-6 contain, apparently, overly optimistic approximations that are probably overestimated by 1.5-2 times. On the other hand, the UCN flux density in the moderator is inversely proportional to the square of the neutron temperature, and the use of a cryogenic moderator can not only compensate for these errors, but also increase the total estimate by at least several times.

It seems that the concept of an intense UCN source based on a pulsed reactor presented here should be the object of more careful calculations, the results of which, in their turn, should be confirmed by experiments. This seems to be especially relevant in connection with the plans to build a new intense neutron source IBR-3 "Neptune" at JINR.

\section*{Acknowledgments}

The authors are grateful to E.V. Lychagin, A.Yu. Muzychka, Yu.N. Pokotilovsky, and V.N. Shvetsov for fruitful discussions.

\bibliographystyle{elsarticle-num} 

\bibliography{UCNsrs_arxiv.bib}
\end{document}